\documentclass[aps,prb,superscriptaddress,amsmath, amssymb,floatfix,notitlepage]{revtex4-1}

\usepackage{graphicx, color, colortbl}
\AtBeginDocument{\hypersetup{colorlinks=true,allcolors=blue}}

\newcommand{\bM}{\mathbf{M}}

\newcommand{\br}{\mathbf{r}}

\renewcommand{\vec}[1]{{{\mathbf{\boldsymbol #1}}}}

\newcommand{\hn}{\hat{n}}

\begin{document}

\hrule
\vspace{1mm}
Author's version of {\it``Shape-dependence of the barrier for skyrmions on a two-lane racetrack''}, Proc. SPIE 9931, Spintronics IX, 993136 (September 26, 2016); doi:10.1117/12.2237624.

Copyright 2016 Society of Photo-Optical Instrumentation Engineers. One print or electronic copy may be made for personal use only. Systematic reproduction and distribution, duplication of any material in this paper for a fee or for commercial purposes, or modification of the content of the paper are prohibited.
\vspace{1mm}
\hrule
\vspace{1cm}

\title{Shape-dependence of the barrier for skyrmions on a two-lane racetrack}
\author{Jan M\"uller}
\email{jmueller@thp.uni-koeln.de}
\affiliation{Institut f\"ur Theoretische Physik, Universit\"at zu K\"oln, D-50937 Cologne, Germany}

\begin{abstract}
Single magnetic skyrmions are localized whirls in the magnetization with an integer winding number. They have been observed on nano-meter scales up to room temperature in multilayer structures. Due to their small size, topological winding number, and their ability to be manipulated by extremely tiny forces, they are often called interesting candidates for future memory devices. 
The two-lane racetrack has to exhibit two lanes that are separated by an energy barrier. The information is then encoded in the position of a skyrmion which is located in one of these close-by lanes. The artificial barrier between the lanes can be created by an additional nanostrip on top of the track.
Here we study the dependence of the potential barrier on the shape of the additional nanostrip, calculating the potentials for a rectangular, triangular, and parabolic cross section, as well as interpolations between the first two. We find that a narrow barrier is always repulsive and that the height of the potential strongly depends on the shape of the nanostrip, whereas the shape of the potential is more universal. We finally show that the shape-dependence is redundant for possible applications.
\end{abstract}

\maketitle

\section{Introduction}

\begin{figure} [b]
 \begin{center}
  \begin{tabular}{c} 
   \includegraphics[width=0.95\textwidth]{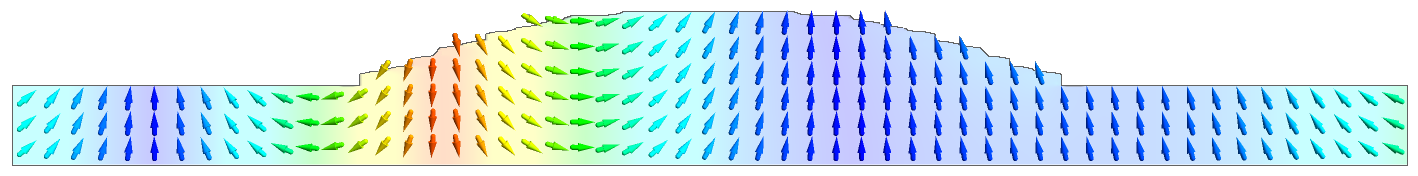}
  \end{tabular}
 \end{center}
 \caption{ \label{fig1} 
 Example of the magnetization (arrows) in a two-lane racetrack with a non-rectangular nanostrip on top. Side view on a cut (y-z-plane) through the two-lane racetrack. The color encodes the z-component of the magnetization. Result from micromagnetic simulation, see Sec.~\ref{sec:model}. The arrows picture the magnetization on a subset of simulated lattice sites.}
\end{figure}

The ever growing demand for more storage capacity on steadily shrinking scales in electronic devices requires information carrying units at minimal size. But even on the smallest scales, the information in these units needs to be controllably read and written, while at the same time it has to be stable against thermal fluctuations.
Particle-like whirls with integer topological winding number in the magnetization, so-called {{\it skyrmions}\cite{lit:muehlbauer,
lit:SkyrmionLatticeRealSpaceObservation,
lit:SkyrmionLatticeAtomicScale}}, were observed on scales down to nanometers and can be created and destroyed on demand\cite{lit:SkyrmionWriteAndDeleteRomming}.
Their efficient coupling to electronic\cite{lit:SkyrmionMotionUltralowCurrent, lit:SkyrmionFlowUltralowCurrent, lit:SkyrmionMotionUniversalCurrentVelocity} or magnonic\cite{lit:SkyrmionLatticeSpinTorquesRotation,lit:SkyrmionMagnonScattering} currents makes ultra-low current densities sufficient to manipulate the motion of skyrmions. Therefore they are often discussed as possible candidates for information carriers in future high-density, non-volatile memory devices\cite{lit:SkyrmionStorageKiselev}, and in particular due to their high mobility in racetrack layouts\cite{lit:SkyrmionRacetrack,lit:SkyrmionTwoLaneRacetrack}.

Chiral skyrmions have been predicted theoretically in chiral magnets\cite{lit:bogdanov,lit:SkyrmionLatticeGroundState} and were experimentally discovered as the skyrmion lattice phase in the chiral magnet MnSi\cite{lit:muehlbauer}.
They are stabilized by Dzyaloshinskii-Moriya interaction (DMI) which favors a twisting in the magnetization and arises due to spin-orbit coupling combined with broken inversion symmetry.
The broken inversion symmetry, however, does not necessarily arise from the broken inversion inside the unit cell as in MnSi. Also interfaces (or surfaces) break inversion symmetry and contribute to an interfacial DMI\cite{lit:SurfaceDMI} (iDMI).
A system with iDMI can be constructed in multilayer systems, where the thermal stability of skyrmion is enhanced by chiral stacking of the multilayers\cite{lit:InterfacialDMIstacking} which lead to the stabilization of single skyrmions at room temperature\cite{lit:SkyrmionsAtRoomTemperature}.
Since then, even the driving of skyrmions by an electric current in a racetrack was shown in experiments\cite{lit:SkyrmionCurrentDrivingExperiment} and for detecting a skyrmion in a future device, their non-coplanar magnetoresistance\cite{lit:SkyrmionElectronicDetection} is a delightful candidate.

As an orthogonal ansatz to the skyrmion racetrack\cite{lit:SkyrmionRacetrack}, which is a nanowire with a sequence of skyrmions in fixed distances, the two-lane skyrmion racetrack\cite{lit:SkyrmionTwoLaneRacetrack} was proposed. Here the information is not encoded in the distance between skyrmions but by the lane on which the skyrmion moves. The two lanes are separated by an energy barrier that is large enough to suppress thermally activated switching of lanes. In micromagnetic simulations the energy barrier was achieved by an additional strip on top of the racetrack, see Fig.~\ref{fig1}, which creates a repulsive potential for skyrmions. To preserve the order of skyrmions, the two lanes are close enough such that the repulsive interaction between skyrmions on different lanes prevents them from overtaking each other. 

A challenging task for the construction of a two-lane racetrack seems to be the preparation of the nanostrip, since the originally proposed geometry demands sharp edges and flat surfaces which in practice is hard to prepare. 
Hence the aim of this paper is to analyze how the energy barrier depends on the shape of the physical barrier and in particular if this concept is robust against deviations from the actual shape.

\section{The Model}
\label{sec:model}

We consider a magnetic multilayer system with interfacial Dzyaloshinskii-Moriya interaction (iDMI) which arises from the broken inversion symmetry at the interfaces of the stacked layers.
The dominant interaction in the system is the ferromagnetic exchange $A$, followed by the much weaker iDMI $D$.
Since the latter interaction prefers the magnetic moments in the system to be pairwise orthogonal, the uniform texture that is preferred by the ferromagnetic exchange as a compromise winds into smoothly twisting spirals.
The wavelength of these spirals, $\lambda=1/Q=2A/D$, defines the length scale on which the magnetization changes.
It ranges from 1-100 nm for different systems with skyrmions\cite{lit:SkyrmionMotionAndPinningByDefectsExperiment,lit:SkyrmionCurrentDrivingExperiment,lit:SkyrmionsAtRoomTemperature}.
Since typically this magnetic length $\lambda$ is much larger than the length scale of the underlying atomic lattice $a$, $\lambda\gg a$, we can apply a continuous theory in which the magnetization is represented by a continuous vector field $\bM(\br)$ with $\|\bM\|=M$.
Furthermore, including an external magnetic field $\mu_0 H$ and uniaxial anisotropy $K$, the free energy functional, $F = \int d^3 \vec r \,\mathcal{F}$, reads
\begin{equation}
\mathcal{F} = A (\partial_\alpha \hat{n}_\beta)^2 + D \left( \hn_i \partial_i \hn_z-\hn_z \partial_i \hn_i   \right) - \mu_0 H M \hn_z - K \hn_z^2 \text{,}
\label{eq:energy}
\end{equation}
with the normalized magnetization $\hn=\bM/M$ and summation indices $\alpha,\beta = x,y,z$ and $i = x,y$.
In the following we will neglect the anisotropy term, $K=0$, since it does not change our results qualitatively and apply a magnetic field to stabilize the polarized phase and single skyrmions therein.
The dynamics of the system (including an applied spin current with drift velocity $\mathbf{v_s}$) are governed by the Landau-Lifshitz-Gilbert (LLG) equation
\begin{equation}
\left[\partial_\text{t} + \left( \mathbf{v}_s \cdot \nabla \right)\right] \hn= -\gamma \hn \times \mathbf{B_\text{eff}} 
+ \alpha \hn \times \left[ \partial_\text{t} \hn + \frac{\beta}{\alpha} \left( \mathbf{v}_s \cdot \nabla \right) \hn \right] \text{,}
\label{eq:LLG}
\end{equation}
with the effective magnetic field $\vec B_{\rm eff} = - \frac{1}{M} \delta F/\delta \hat n$ and the spin density $s=\frac{M}{\gamma}$.
The phenomenologic adiabatic and non-adiabatic damping coefficients $\alpha$ and $\beta$ are chosen equally $\alpha=\beta=0.1$ throughout this paper.
Since the continuous field theory can be rescaled, the units we use in the following are in accordeance with previous works the 
dimensionless momentum $Q = \frac{D}{2A}$, 
total energy $E_D = \frac{2 A}{Q}$, 
magnetic field $h_D = \frac{2 A Q^2}{\mu_0 M}$,
time $t_D = \frac{s}{2AQ^2}$, 
and velocity $v_D = \frac{2AQ}{s}$.
In the whole paper, the magnetic field is set to $h=0.75 h_D$.

\begin{figure} [t]
 \begin{center}
  \begin{tabular}{c} 
   \includegraphics[width=0.45\textwidth]{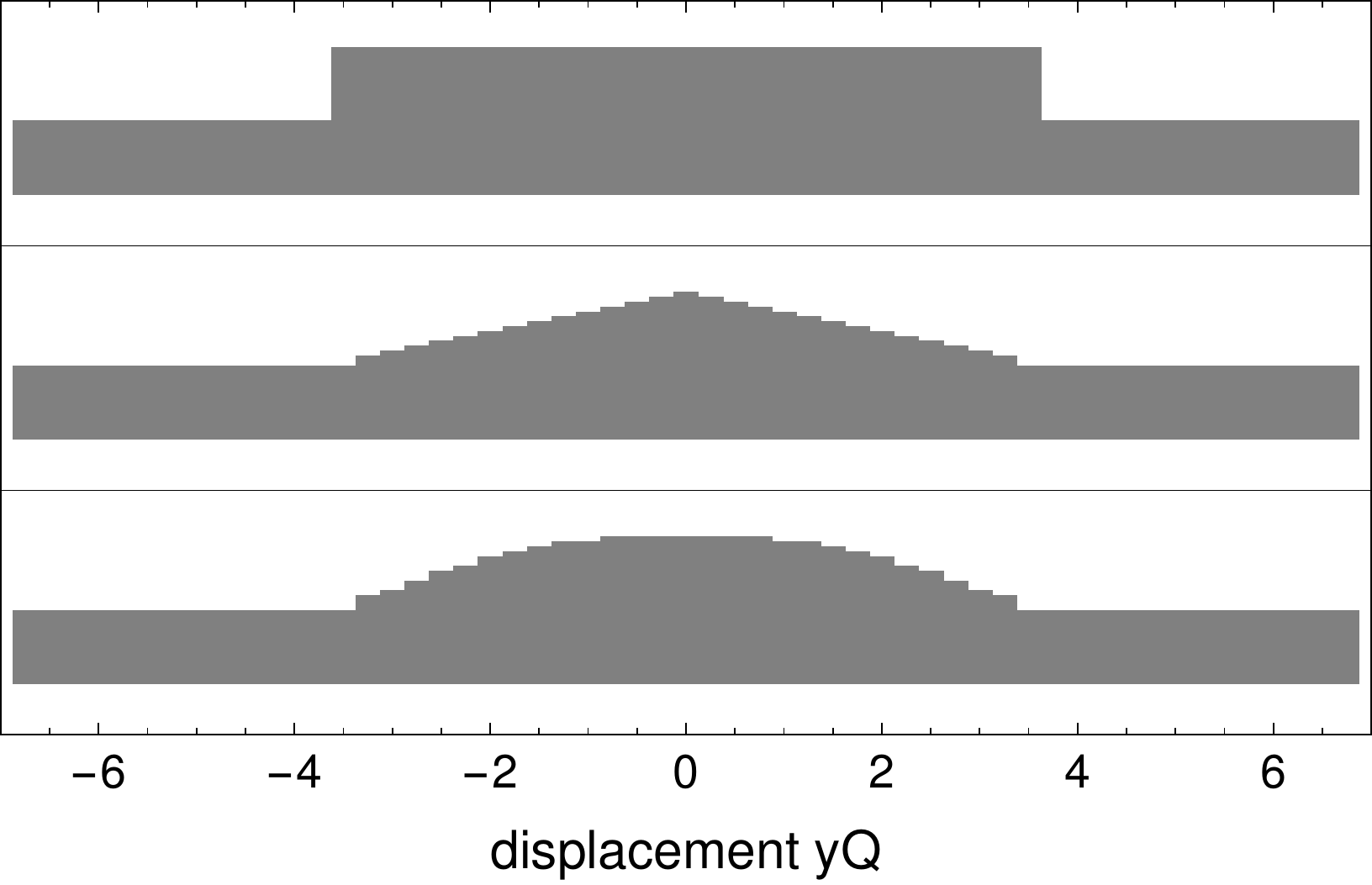}
   \hspace{0.05\textwidth}
   \includegraphics[width=0.45\textwidth]{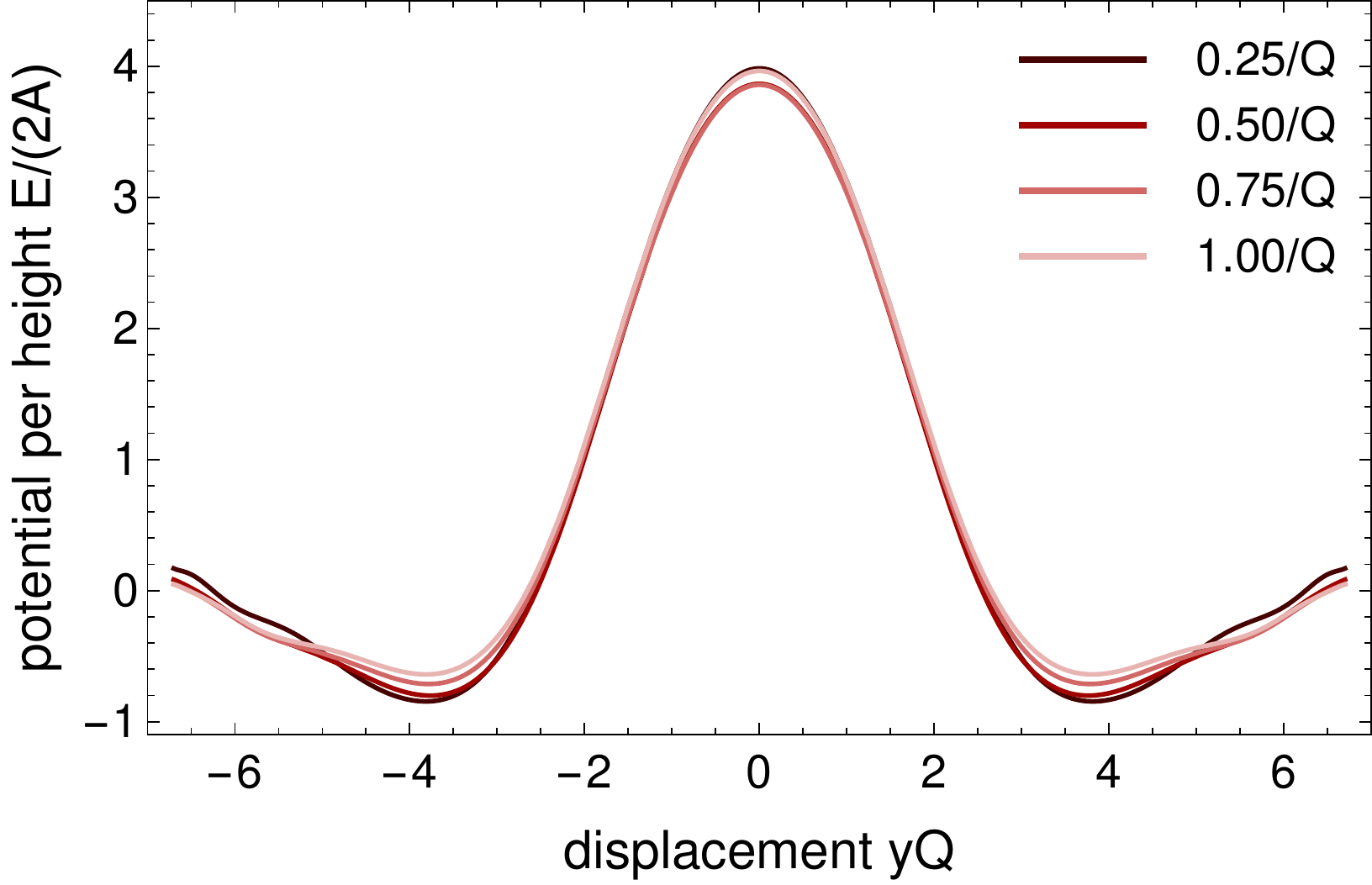}
  \end{tabular}
 \end{center}
 \caption{ \label{fig2} 
 Left: The shapes of the nanostructures on top of the racetrack ({\it rectangular}, {\it triangular}, and {\it parabolic}) that are considered in this paper, here all shown at width $7.25/Q$. 
 Right: Linear scaling of the potentials with the height of the structure. Plot shows the potential per height calculated for the triangular barrier with width $6.25/Q$  and various heights as shown in the inset.}
\end{figure}

For the evaluation of the different potentials that are discussed in this work we use the same methods as in Ref.~\citenum{lit:SkyrmionTwoLaneRacetrack}.
However, in order to be able to resolve the nanostrip sufficiently detailed for an analysis of the shape-dependence, we choose the numeric lattice discretization equal in the plane of the track, $a_x=a_y=0.25/Q$, and finer in the z-direction perpendicular to the track, $a_z=0.05/Q$.
With these discretizations we simulate a single skyrmion in a two-lane racetrack with unpatterned bottom layers of width $13.75/Q$ and height $0.75/Q$ and on top of that {\it rectangluar}, {\it triangluar}, or {\it parabolic} barriers of height $0.75/Q$ and various widths, see Fig.~\ref{fig2} left.
Generally, the geometry has more tunable parameters, but it has been shown in Ref.~\citenum{lit:SkyrmionTwoLaneRacetrack} that this width of the bottom layers works well for a two-lane racetrack. The heights of the bottom and top part of the nanostructure could be varied in the simulations, but it has furthermore been shown that the potential obeys a scaling law, see Fig.~\ref{fig2} right. Due to the extreme robustness of the magnetic structure against deformations along the z-direction, the total potential can be expressed as a superposition of the potential from the bottom layers and the top layers, each scaled with their respective heights:
\begin{equation}
V(y) = h_{\text{track}} v^\text{track}(y) + h_{\text{barrier}} v^\text{barrier}(y) \text{.}
\label{eq:potential_superposition}
\end{equation}
Since we want to analyze the impact of the shape of the top structure on the potential, we will in the following exclusively focus on the potential per height of this upper layers, $v^\text{barrier}(y)$.

\section{Barriers and potentials}
\label{sec:potentials}

\subsection{Comparison of the different barriers}
\label{sec:potentials-compare}

\begin{figure} [t]
 \begin{center}
   \includegraphics[width=0.45\textwidth]{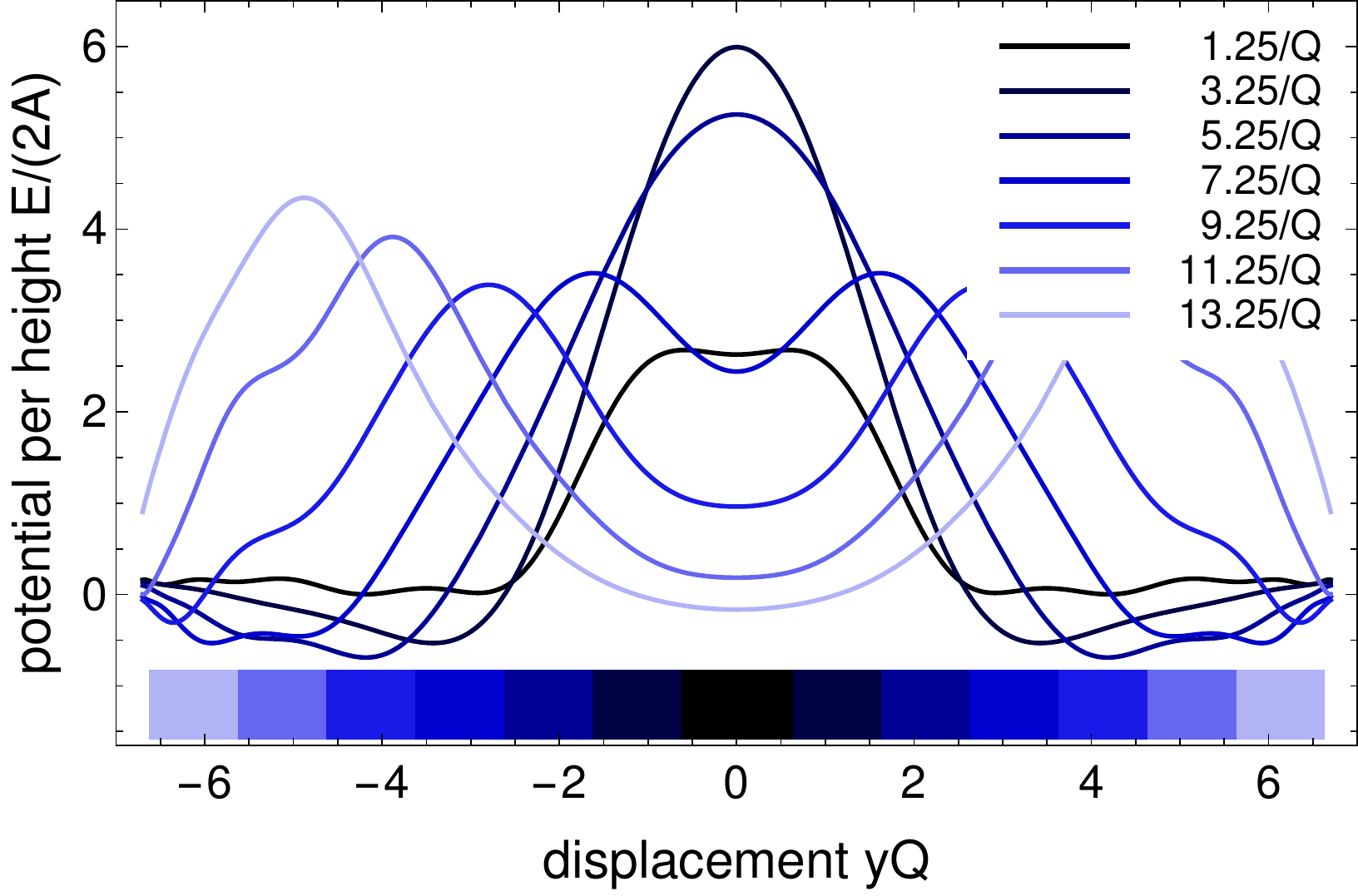}
   \hspace{0.05\textwidth}
   \vspace{0.03\textwidth}
   \includegraphics[width=0.45\textwidth]{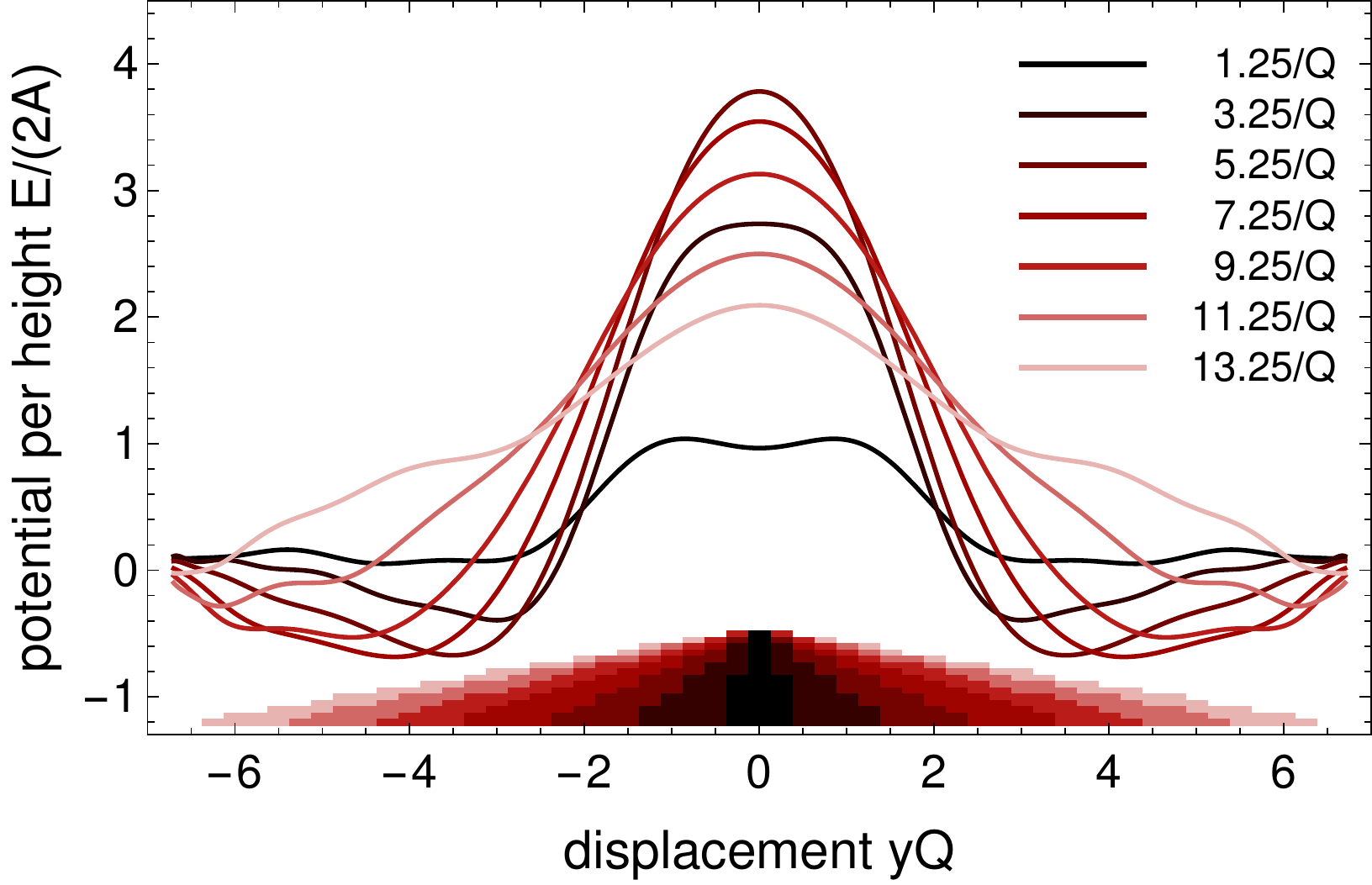}
   \includegraphics[width=0.45\textwidth]{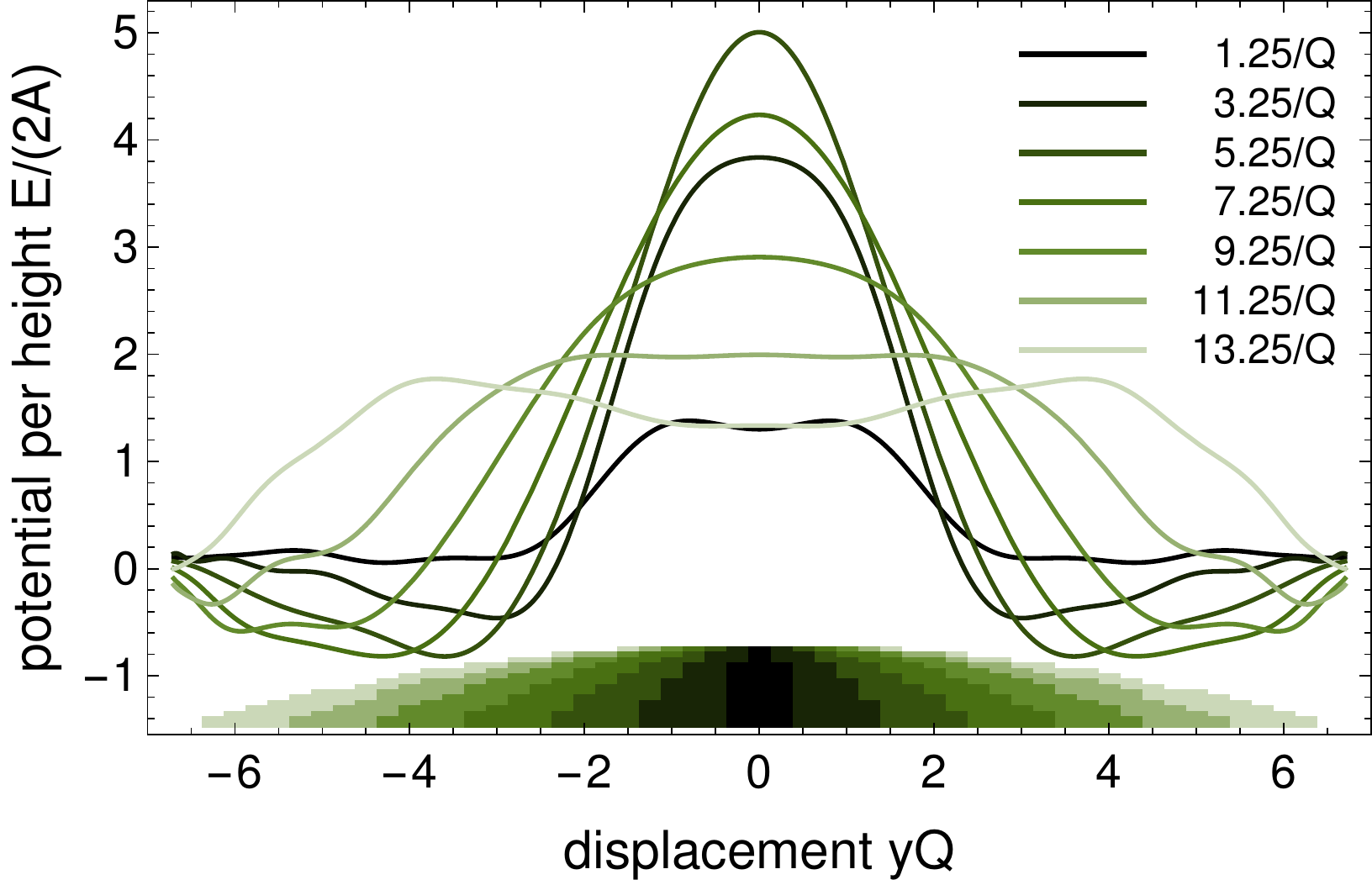}
   \hspace{0.05\textwidth}
   \includegraphics[width=0.45\textwidth]{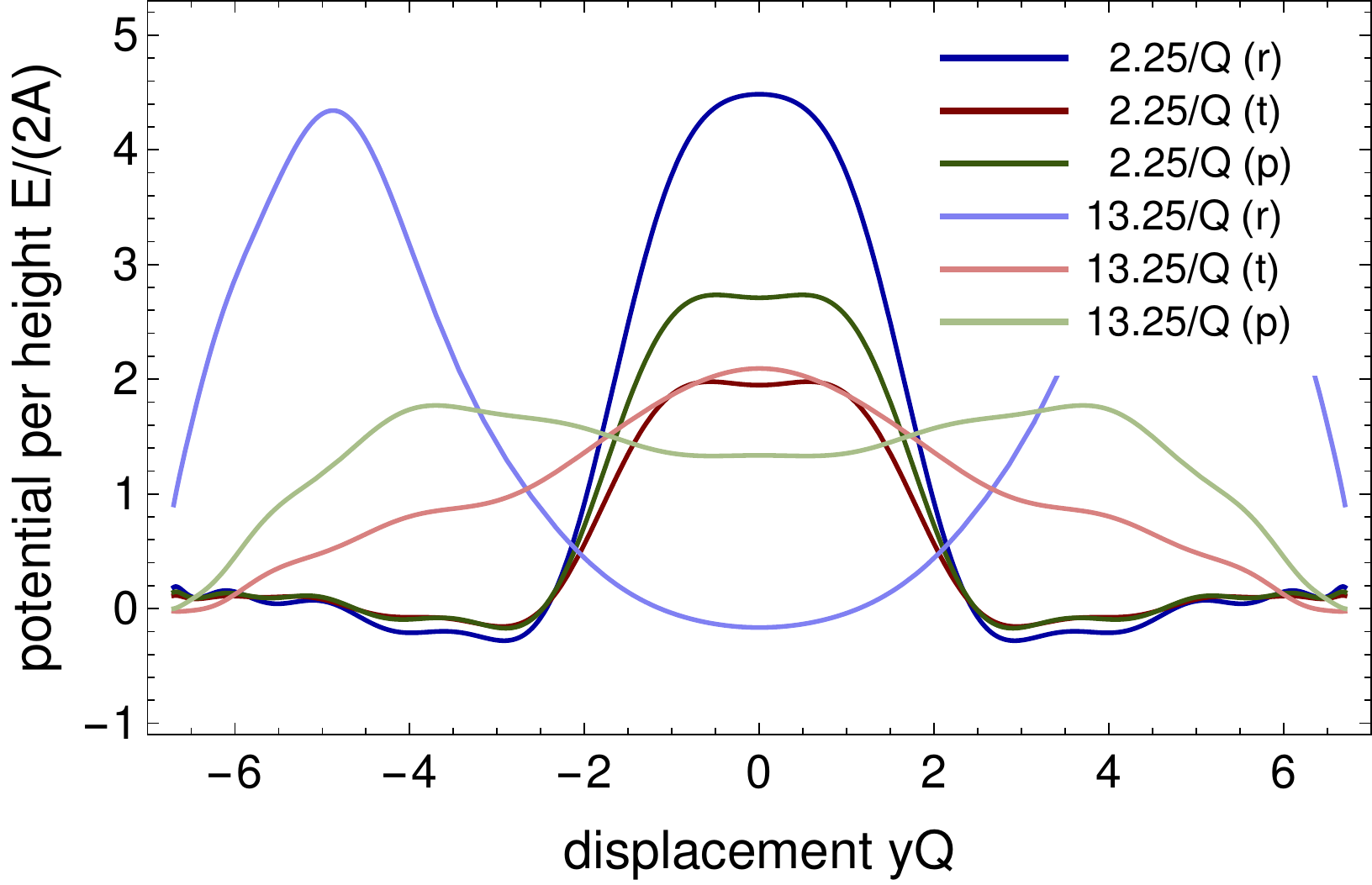}
 \end{center}
 \caption{ \label{fig3} 
 Potentials per height for different shapes of nanostrips on the racetrack.
 The color (blue, red, green) denotes the shape (rectangular, triangular, parabolic).
 The brightness (dark to light) corresponds to the width of the strip (narrow to broad) as more precisely given in the insets.
 The solid blocks on the bottom of the first three figures show, as a guide to the eye, the corresponding cross section of the strip.
 }
\end{figure} 

The potentials per height, $v^\text{barrier}(y)$, for all three considered shapes of nanostrips (rectangular, triangular, parabolic) are shown in the first three plots in Fig.~\ref{fig3} for the whole range of possible widths, i.e. ranging from a narrow peak with a width of only three numeric discretization lengths (black curves) to a broad shape which covers the whole racetrack underneath (very light curves). The potential per height of the bottom layers of the track, $v^\text{track}(y)$, is not shown explicitly but is approximately given by the potential of the broadest rectangular strip.

The exact dependence of the potential on the geometry of the strip is for all three cases very complex. For low widths, which were considered the important regime for a two-lane potential, a repulsive peak grows with the width in the center of the track, independent of the shape of the nanostrip. The height of the peak, however, differs for the various cases, see Fig.~\ref{fig3} bottom right, up to a factor two: In the narrow peak is not much room for the magnetization to develop structures that adapt to the shape. This effect can more likely be explained by the fact that the volume of the triangular nanostrip is twice as low as in the rectangular case. Consequently, for a two-lane racetrack any of the different shapes can in principle be considered as suitable for producing the desired energy barrier, if the heights are scaled properly.

If the nanostrip is about half the width of the total racetrack, the different shapes start to develop a more characteristic potential per height, $v^\text{barrier}(y)$, see Fig.~\ref{fig3} bottom right.
In the case of the rectangular nanostrip, the repulsive peak in the center decays into two peaks which with growing width move apart to the edges of the racetrack. The result is a local minimum in the center of the track which grows in both, depth and width. In the limit where the width of the nanostrip is equal to the width of the racetrack the potential minimum in the center is the reason why skyrmions in a racetrack are repelled from the edges and it becomes even the global minimum, i.e. the racetrack prefers to have skyrmions for this choice of magnetic field\cite{lit:SkyrmionTwoLaneRacetrack}, $h=0.75h_D$. 

Differently, the center of the triangular geometry, where the only edge in the continuous limit is located, remains repulsive for all widths and the peak in the center is always sharper than the minimum of the potential of the bottom layers of the track. For the construction of a two-lane racetrack this means in particular that triangular nanostrips of any width can in total produce a two-lane potential if the thickness of underlying bottom layers is chosen accordingly.

The parabolic shaped nanostrip with its flat center and curved edges acts as a compromise between the flat rectangular shape and the triangular with its constant gradient in height. For larger widths the peak in the center decays just as in the triangular case without the formation of a minimum (except for the limit of same width as the racetrack), but in contrast to the sharp peak of the triangular case it becomes much broader, comparable to the two disconnecting peaks of the rectangular setup. Since the repulsive peak is extremely flat around the center, the parabolic nanostrip can only up to limited width be able to yield a two-lane potential. Interestingly, in the limit of equal width with the racetrack, the parabolic nanostrip exhibits a minimum in the center and therefore keeps skyrmions naturally on the (single-lane) racetrack. The fact that this minimum is only very flat, on the other hand, is an indicator that a normal racetrack with an approximately parabolic cross section can hardly keep the 
skyrmions on the track. In conclusion we find for the cross section of the unpatterned racetrack layers in the bottom half as for the nanostrip on top that the sharper the edges are, the stronger is the repulsion.

\subsection{Smeared out edges}
\label{sec:potentials-washout}

\begin{figure} [t]
 \begin{center}
   \includegraphics[width=0.45\textwidth]{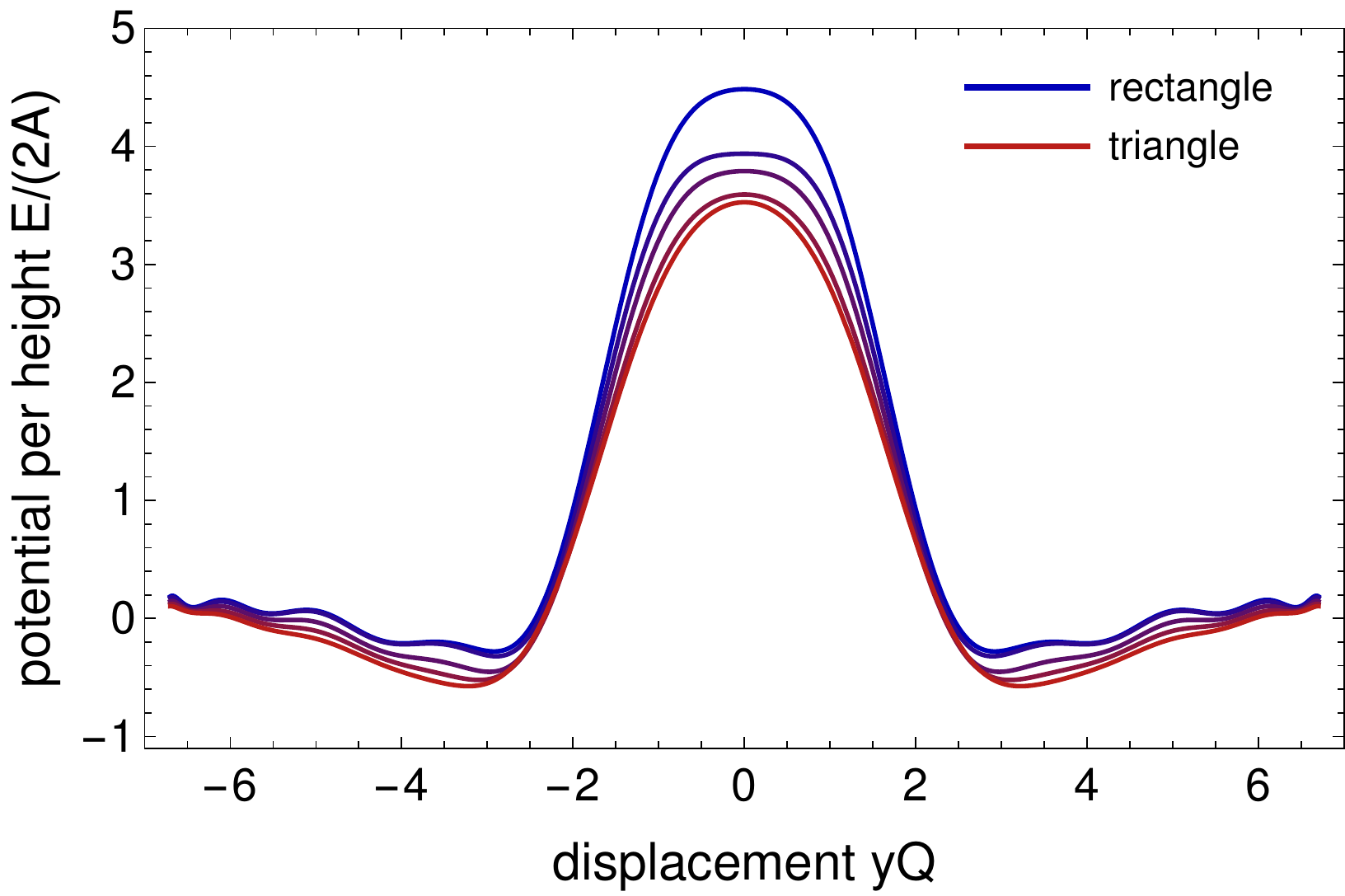}
   \hspace{0.05\textwidth}
   \includegraphics[width=0.194362\textwidth]{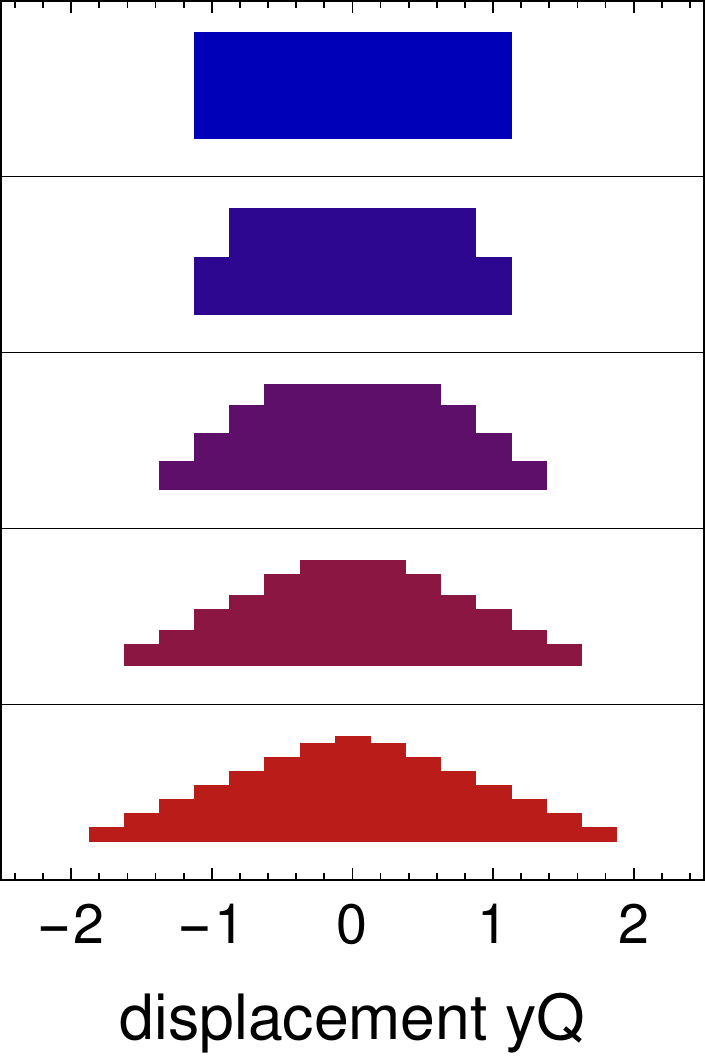}
 \end{center}
 \caption{ \label{fig4} 
 Left: Potentials per height for different shapes of nanostrips on the racetrack.
 The color (blue to red) denotes the shape (rectangular to triangular) and corresponds to the shapes shown in the figure on the right.
 Right: Cross sections of the different nanostrips used for the potentials in the figure on the left.
 }
\end{figure}

For the realization of a two-lane racetrack, a sharp repulsive potential peak is needed and can, according to the previous section, be constructed from a nanostrip of any shape if it is narrow enough.
In the following we therefore discuss shortly the effect of an imperfect preparation method, which is supposed to prepare the rectangular nanostrip discussed in Ref.~\citenum{lit:SkyrmionTwoLaneRacetrack} ($2.25/Q$ width, $0.75/Q$ height) but does not resolve the edges sharply such that they smear out.
In our simulations, we smear out the original shape by replacing the sharp edges by edges with a linear slope. Upon increasing the slope, the rectangle is smoothly (within the limits of the numerical discretization) transformed into a trapezoid and finally a triangle of doubled width, see Fig.~\ref{fig4} right. The respective potentials are shown on the left hand side of Fig.~\ref{fig4}.
The results are consistent with the above claim, namely that the potential for narrow nanostrips is only slightly affected by the exact shape whereas the amplitude depends to good approximation linearly on the volume of the nanostrip. Hence the potentials are for all shapes qualitatively similar and differ only slightly in height, which can be explained with slightly lower volumes for the more triangular shapes due to numerical discretization.
In conclusion we find that the actual shape of the nanostrip is of minor importance if the volume of the nanostrip is conserved.

\section{Summary}

The two-lane skyrmion racetrack, with its new concept of encoding information in a sequence of repulsive skyrmions on different lanes of a racetrack, is a promising new approach towards skyrmion-based storage devices.
The two lanes can be created by a repulsive potential in the center of a racetrack.
One can achieve such a potential by an additional nanostrip on top of the racetrack. We show that the repulsive potential generally depends on the shape of this nanostrip, especially if the nanostrip is wide. However, various shapes fulfill the requirements for the potential barrier if they are sufficiently narrow. We show that for a particular example, that has been proposed as being suitable for practical purposes, the exact shape of this nanostrip is only of minor importance. 
We therefore believe that the separation of lanes by a nanostrip is a robust method since it is mostly unaffected by small deformations of the strip.

\acknowledgments
This work was supported by Deutsche Telekom Stiftung and the Bonn-Cologne Graduate School of Physics and Astronomy BCGS.
We furthermore thank the Regional Computing Center of the University of Cologne (RRZK) for providing computing time on the DFG-funded High Performance Computing (HPC) system CHEOPS as well as support.


%

\end{document}